\documentclass[aps,prl,preprint,superscriptaddress,showpacs]{revtex4}
\usepackage{setspace}
\usepackage{amsmath}
\usepackage{amssymb}
\usepackage{epsfig}
 \normalsize
\begin{document}
\preprint{}
\title{Nematic phase transitions in mixtures of thin and thick colloidal rods}
\author{Kirstin Purdy}
\affiliation{Martin Fisher School of Physics, Brandeis University,
Waltham, Massachusetts 02454}
\author{Szabolcs Varga}
\affiliation{Department of Physics, University of Veszpr\'{e}m,
H-8201 Veszpr\'{e}m, PO Box 158, Hungary}
\author{Amparo Galindo}
\author{George Jackson}
\affiliation{Department of Chemical Engineering and Chemical
Technology, South Kensington Campus,
 Imperial College London, London SW7 2AZ, UK}
\author{Seth Fraden}
\affiliation{Martin Fisher School of Physics, Brandeis University,
Waltham, Massachusetts 02454}
\date{\today}
% ----------------------------------------------------------------
\begin{abstract}
We report experimental measurements of the phase behavior of
mixtures of thin (charged semiflexible {\it fd} virus) and thick
({\it fd}-PEG created by covalently grafting poly-(ethylene
glycol) to the surface of {\it fd} virus) rods. The {\it fd}-PEG
are sterically stabilized but the {\it fd} virus are charged, thus
varying the ionic strength of a mixture of {\it fd} and {\it
fd}-PEG varies the effective diameter of the bare {\it fd} rods,
and the effective diameter ratio ($d\equiv
D_{\mbox{\scriptsize{{\it fd}-PEG}}}/D_{\mbox{\scriptsize{{\it
fd}}}}$). We examine the phase diagrams of the rod mixtures and
find isotropic-nematic, isotropic-nematic-nematic and
nematic-nematic coexisting phases with increasing concentration.
In stark contrast to predictions from earlier theoretical work, we
observe a nematic-nematic coexistence region bound by a lower
critical point. Moreover, we show that a rescaled Onsager-type
theory for binary hard rod mixtures qualitatively describes the
observed phase behavior.
%can cut this much shorter

\end{abstract}
 \maketitle

The entropy driven phase transition of monodisperse suspensions of
purely repulsive rods from an isotropic to an aligned nematic
phase has been extensively studied experimentally
\cite{Fraden95,Sato96}, theoretically \cite{Onsager49,Vroege92},
and computationally \cite{Bolhuis97a}. Binary mixtures of hard
particles of different aspect ratios and/or shape can exhibit a
much richer phase behavior. Quantitatively studying a binary
hard-rod mixture is the first step towards understanding the
impact of size polydispersity on the phase separation of
concentrated suspensions of rodlike macromolecules. Length and
diameter polydispersity are prevalent in suspensions of many
synthetic and biological rodlike particles including F-actin,
microtubules, and DNA. Theoretical studies of binary hard-rod
mixtures predict that in addition to isotropic-nematic (I-N)
coexistence, isotropic-nematic-nematic (I-N-N),
isotropic-isotropic (I-I), and nematic-nematic (N-N) coexistence
are possible when the length or diameter ratio of the particles is
large enough \cite{Abe78,Lekkerkerker84,Vroege93,vanRoij98,
Hemmer99,Speranza02,Varga03}. However, the topology of the
theoretical phase diagram of binary mixtures of rods, including
the progression of the phase behavior from a totally miscible
nematic to a demixed N-N state and the related existence of N-N
critical points, is a subject of considerable debate as the
predicted phase behavior of the more concentrated phases (N-N and
I-N-N) is extremely sensitive to the approximations employed
\cite{Vroege93,vanRoij96b,vanRoij98,Varga00,Varga03}. The focus of
past experimental studies has been on binary mixtures of rods of
varying length where I-N, I-N-N and N-N coexistence have been
measured \cite{Itou84, Kajiwara86,Lekkerkerker95,Sato96}. Because
of polydispersity in the particle size, high solution viscosity
and/or weak attractions these measurements were constrained to low
concentrations near the I-N transition
\cite{Itou84,Lekkerkerker95,Sato96}. As a consequence,
experimental studies of  N-N phase behavior are rare. In this
Letter, we present experimental measurements of the phase behavior
of binary mixtures of rods of varying \emph{diameter} ratio and
equal length to very high (nematic) concentrations. These rods are
strongly repulsive and highly monodisperse. By examining the phase
diagrams well into the nematic region, we find a lower critical
point in the N-N coexistence where previously only an upper
critical point had been predicted. We compare these unexpected
results to the predictions of a scaled Onsager theory
\cite{Varga03}.

The theory describing the liquid crystal phase behavior of binary
rod mixtures is closely related to the theory of orientational
ordering developed by Onsager for monodisperse rods
\cite{Onsager49}. Onsager showed that the decrease in
orientational entropy accompanying the formation of an aligned
nematic phase is more than compensated for by an increase in the
free volume entropy of the rods. The I-N phase transition
predicted for monodisperse rods is in excellent agreement with
experiments \cite{Fraden95}. By extending Onsager's theory for the
free energy of hard rigid rods in the limit of the second virial
coefficient (second virial theory, or SVT) to binary mixtures
\cite{Onsager49}, the phase behavior has been determined for
binary mixtures of rods of different diameter \cite{vanRoij98},
and different length \cite{Lekkerkerker84,Vroege93,Varga00}.
Onsager's second virial expansion of the free energy is, however,
quantitatively accurate only for dilute suspensions of particles
of large aspect ratio. For concentrated phase transitions (such as
a N-N transition), both ``end-effects" due to the finite size of
the rods (typically neglected for Onsager-limit studies) and
higher virial coefficients are significant, even in the limit of
large aspect ratio. Therefore, to study the phase behavior of
binary mixtures of rods of short, experimentally accessible,
aspect ratios and to describe the phase behavior at high
concentrations, we adopt a Parsons-Lee (P-L) free energy
\cite{Parsons79,Lee87} as applied to mixtures of rigid rods by
Varga {\it et al.}~\cite{Varga03}. The P-L approach approximates
higher virial coefficients by interpolating between the
Carnahan-Starling free energy for hard spheres and the Onsager
free energy for long hard rods. The free energy is described by:
\begin{eqnarray}
\frac{\beta F}{N}=\mbox{const}+\sum_{i=1}^2 x_i(\ln(x_i \rho)
+\sigma(f_i)) \nonumber + \frac{\beta F^{\mbox{\scriptsize
HS}}_{\mbox{\scriptsize exc}}}{N 8 v_{\mbox{\scriptsize HS}}}
\sum_{i=1}^2\sum_{j=1}^2 \langle v_{ij}\rangle
\end{eqnarray}
Here $\rho=(N_1+N_2)/V$ is the total number density, $x_i$ and
$f_i(\Omega)$ are the mole fraction and orientational distribution
function of rods of type $i$, respectively, and $\sigma$ is the
single particle orientational entropy \cite{Varga03}. In the
second term, $F^{\mbox{\scriptsize HS}}_{\mbox{\scriptsize exc}}$
is the residual free energy of a system of equivalent hard spheres
of volume $v_{\mbox{\scriptsize HS}}=(x_1 v_1+x_2 v_2)$
\cite{Varga03} , and $v_i$ is the volume of rod $i$. Furthermore,
$\langle v_{ij}\rangle=\int\int x_i x_j
v_{ij}(\Omega_1,\Omega_2)f_i(\Omega_1) f_j(\Omega_2)
d\Omega_1d\Omega_2$ is the full orientationally averaged excluded
volume ($v_{ij}$) between rods $i$ and $j$, with $\Omega$ being
the orientational unit vector (polar and azimuthal angles). The
end-effects are included the excluded volume term\cite{Varga03}.
The theoretical phase diagrams presented here are calculated
numerically with this free energy functional using techniques
previously described \cite{Varga03}.

%\section{Experimental System}
Mixtures of two well characterized systems: the charged
semiflexible {\it fd} virus \cite{Fraden95} and {\it fd} virus %\cite{Tang95}
irreversibly coated with the neutral polymer poly(ethylene glycol)
(PEG) \cite{Dogic01,Grelet03} are investigated experimentally. The
bare {\it fd} virus will serve as our thin rod, while the polymer
coated {\it fd} ({\it fd}-PEG) will be our thick rod. The {\it fd}
virus is grown and purified as previously described
\cite{Maniatis89b}, and it is characterized by its length $L=880$
nm, diameter $D=6.6$ nm, molecular weight $M_w=1.64\times 10^7$
g/mol, and persistence length, defined as the length over which
the tangent vectors along a polymer are correlated, of $P=$2.2
$\mu$m \cite{Fraden95}. The {\it fd} rods also have a linear
surface charge of 10 e$^-$/nm at pH 8.2, thus to compare the phase
behavior of {\it fd} with predictions for hard rods the
electrostatic interactions are incorporated into an effective
diameter $D_{\mbox{\scriptsize{eff}}}$ which is larger than the
bare diameter and increases with decreasing ionic strength
\cite{Onsager49, Stroobants86a}. Over a wide range of ionic
strengths, monodisperse suspensions of {\it fd} are known to
undergo an isotropic to cholesteric (chiral nematic) phase
transition which is accurately described by SVT for semiflexible
hard rods of diameter $D_{\mbox{\scriptsize{eff}}}$
\cite{Fraden95}. Because the free energy difference between the
cholesteric and nematic phases is small, comparison with theories
developed for nematic phases is appropriate \cite{deGennes93}.

\begin{figure}
\centerline{\epsfig{file=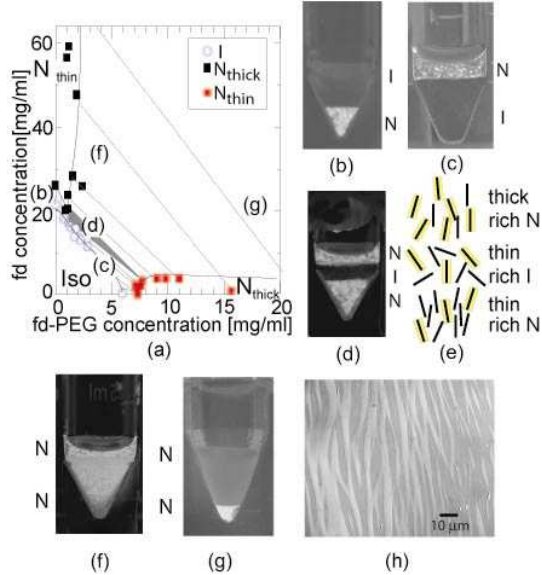,width=2.8
in}}\caption[]{\label{bulk2.fig}(Color Online) Phase separation of
mixtures of {\it fd} and {\it fd}-PEG (20,000 M$_w$ PEG) at 110 mM
ionic strength ($d=3.7$), pH 8.2. The phase diagram is presented
in (a). The dark curves are guides to the eye representing the
phase boundaries of the two nematic phases and the isotropic
phase. The dashed lines indicate coexisting samples, and the three
phase region is indicated by the grey triangle. The coexisting
phases labeled in (a) as viewed under crossed polarizers are: (b)
I-N coexistence; (c) highly fractionated I-N coexistence; (d)
I-N-N three phase coexistence with a schematic representation of
the partitioning of thick and thin rods shown in (e); (f) N-N
demixing just above the triphasic region; and (g) highly
concentrated N-N coexistence showing strong partitioning of the
thick (yellow, {\it fd}-PEG) and thin (white, {\it fd}) rods. The
isotropic phase appears dark and the nematic phase appears light
due to its birefringence. The yellow color is due to the dye on
the {\it fd}-PEG rods. (h) Differential interference contrast
microscopy image of low density {\it fd}-PEG-rich nematic tactoids
in coexistence with high density {\it fd}-rich nematic bulk
phase.}
\end{figure}

The thick rods are formed by attaching an amino-reactive PEG,
$M_w=20 000$ g/mol (SSA-PEG20K, Shearwater Polymer Corp.), to the
exposed amino termini of the viral coat proteins. The dense
polymer coating of approximately 200$\pm$30 PEG20K molecules per
{\it fd} \cite{Grelet03} acts as a steric stabilizer above an
ionic strength of 2 mM. In this regime the isotropic-nematic
(cholesteric) phase transition of monodisperse suspensions of {\it
fd}-PEG is independent of ionic strength \cite{Dogic01}. The
diameter $D_{\mbox{\scriptsize thick}}$ of the thick
(virus+polymer) {\it fd}-PEG rods is thus calculated from the
monodisperse isotropic coexistence concentration ($c_i$) using the
theoretical relationship
$c_{\mbox{\scriptsize{i}}}$[mg/ml]$=222/D_{\mbox{\scriptsize
thick}}$[nm], as calculated from SVT adapted for semiflexible rods
\cite{Chen93, Dogic01}. $D_{\mbox{\scriptsize thick}}$ varies from
25-40 nm depending on the {\it fd}/SSA-PEG20K reaction conditions.
Rods with different $D_{\mbox{\scriptsize thick}}$ were used
separately.

%\begin{figure}\centerline{\epsfig{file=fdpegphasediagram.eps,width=8cm}}
%\caption[Isotropic-nematic transition for suspensions of {\it fd}
%and {\it fd}-PEG20K. ]{\label{fd-pegsketch.fig} (Color Online)
%Isotropic-nematic phase boundary as a function of ionic strength
%for pure {\it fd} (squares) and pure {\it fd}-PEG (20,000 M$_w$
%PEG)(triangles). The data is taken from previous work
%\cite{Tang95, Dogic01}. The theoretical relationship between the
%I-N coexistence concentrations ($c_i$) and the effective diameter
%$c_{\mbox{\scriptsize{i}}}$[mg/ml]$=222/D_{\mbox{\scriptsize
%eff}}$[nm] is shown by the continuous curve, as calculated from
%Onsager's theory for semiflexible rods with $L/P=0.4$
%\cite{Chen93, Dogic01}. The measured concentration is the
%concentration of  bare {\it fd}; the mass of the PEG is not
%considered. In the illustrations we show end-on pictures of
%(a){\it fd} at high ionic strength,(b) {\it fd} at low ionic
%strength and (c) {\it fd}-PEG. The inset illustrates the distance
%of closest approach for (d) two {\it fd} rods, (e) two {\it
%fd}-PEG rods, and (f) {\it fd} and {\it fd}-PEG. The interactions
%are seen to be additive in (d) and (e), non-additive in (f).}
%\end{figure}

%\section{Observation of bulk phase separation}
Samples are prepared at multiple virus compositions and
concentrations to determine the phase behavior of mixtures of {\it
fd} and {\it fd}-PEG. The virus suspensions are dialyzed against
20 mM Tris-HCl buffer at pH 8.15 with NaCl added to change the
ionic strength such that the diameter ratio $D_{\mbox{\scriptsize
thick}}/D_{\mbox{\scriptsize thin}}\equiv d$, where
$D_{\mbox{\scriptsize thin}}=D_{\mbox{\scriptsize eff}}$, varied
from $1.1<d<3.7$. The observed phase separation includes either an
isotropic phase (I) coexisting with a nematic (N) phase, I-N-N
three-phase coexistence, or N-N coexistence. The coexisting
phases, as viewed under crossed polarizers, are depicted in Fig.
\ref{bulk2.fig}. This confirms the theoretical predictions for the
stable coexisting phases for such a system \cite{vanRoij98,
Varga03}.
Phase separation between two isotropic phases (I-I) is not %can cut
observed, presumably because $d$ is not large enough. In Figs.
(\ref{bulk2.fig}c) and (\ref{bulk2.fig}d) the {\it fd}-PEG-rich
nematic phase can be seen to float above the {\it fd}-rich
isotropic phase. Even though the volume fraction of rods is higher
in the nematic phase, the mass density of the {\it fd} rich
isotropic phase is greater than that of the {\it fd}-PEG rich
nematic phase. The mass density difference arises in part because
of the difference in single particle densities $\rho=1.35$ g/ml
for {\it fd} and $\rho=1.007$ g/ml for {\it fd}-PEG. After
equilibration, the concentrations of {\it fd} and {\it fd}-PEG in
the coexisting phases are measured by absorption
spectrophotometry. The optical density ($A$) of {\it fd} is
$A_{3.84 \mbox{\scriptsize ml/mg}}^{269 \mbox{\scriptsize nm}}$
for samples 1 cm thick. To independently measure the
concentrations of thin and thick rods in the coexisting phases,
{\it fd}-PEG is also labeled with the fluorescent dye fluorescein
isothiocyanate (FITC) which has an optical density of $A_{68000
\mbox{\scriptsize L/mol}}^{495 {\mbox{\scriptsize nm}}}$. Before
bulk separation of coexisting phases occurs, phase separation on a
micron length scale can easily be seen using fluorescence and/or
polarization microscopy as shown in Fig. \ref{bulk2.fig}h.

\begin{figure}
\centerline{\epsfig{file=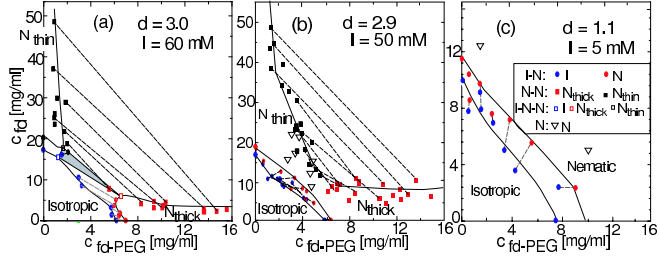, width=3.4 in}}
\caption[]{\label{4diagrams.fig}(Color Online) Phase diagrams of
suspensions of {\it fd}-PEG (20,000 M$_w$ PEG) mixed with {\it fd}
at (a) 60 mM, (b) 50 mM, (c) 5 mM ionic strength, pH 8.2. The
legend for all three diagrams is in (c). The dark curves are
guides to the eye representing the phase boundaries of the two
nematic phases and the isotropic phase. The dashed lines indicate
coexisting samples, and open triangles are single phase nematic
samples as determined by fluorescence microscopy. The three phase
region is indicated by the grey triangle.}
\end{figure}

In Fig. \ref{4diagrams.fig} we present the experimental phase
diagrams of the binary mixtures of {\it fd} and {\it fd}-PEG at
three ionic strengths; corresponding to diameter ratio $d$ ranging
from $1.1<d<3.0$. For large diameter ratios, $d>3.0$, marked
partitioning is seen in the I-N coexistence region, as indicated
by tie lines in the I-N coexistence region which do not pass
through the origin. With increasing overall concentration, a
triangular I-N-N three phase region and a N-N coexistence region
are found (Fig. \ref{4diagrams.fig}a). Almost complete
partitioning of the thick ({\it fd}-PEG) and thin ({\it fd}) rods
is observed in the N-N coexistence region. As $d$ decreases, the
I-N-N coexistence region becomes smaller, and below about $d=3.0$
the triphasic region vanishes. Between $d\sim3.0$ and $d \sim 2.0$
the N-N coexistence is still present at high concentrations, even
in the absence of a well defined I-N-N triangle, suggesting a N-N
region bounded by a lower critical point. Partitioning of the {\it
fd} and {\it fd}-PEG between isotropic and nematic phases also
decreases in the absence of an I-N-N region, as indicated by the
shortening of the I-N tie lines which do not radiate from the
origin from Fig. \ref{4diagrams.fig}b to \ref{4diagrams.fig}c. For
diameter ratios below $d \sim 2.0$ only a single nematic phase is
observed.

\begin{figure}\centerline{\epsfig{file=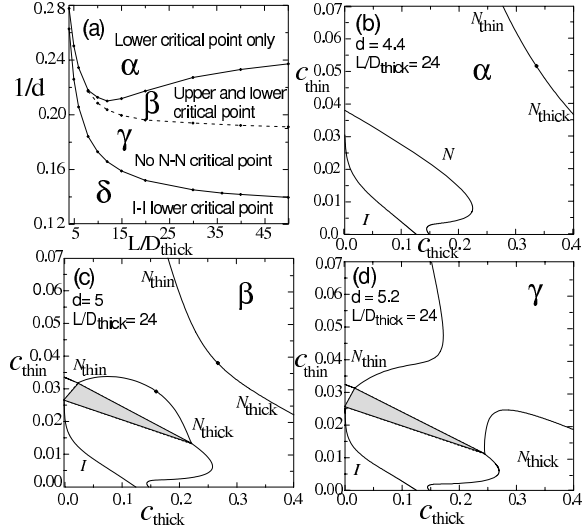,
width=3.in}} \caption[]{\label{4theory.fig} (a) Characterization
of the binary phase behavior of rods of equal length obtained with
P-L scaling of the Onsager free energy as a function of the
inverse diameter ratio $1/d$ and $L/D_{\mbox{\scriptsize thick}}$.
I-N-N coexistence is found below the uppermost line (regions
$\beta$ and $\gamma$). Coalescence of the upper and lower N-N
critical points occurs at the dashed line (region $\gamma$), and
I-I-N coexistence becomes stable below the lower line (region
$\delta$). For $L/D_{\mbox{\scriptsize thick}}<7$, the lower
critical point merges with the I-N coexistence region at the
uppermost solid line. The phase diagrams are calculated for
$L/D_{\mbox{\scriptsize thick}}=24$ with $d=4.4$ (b), $d=5$ (c),
and $d=5.2$ (d). Phase diagrams are presented as a function of
reduced concentration $c_i=v_i N_i/V$. Three phase coexistence is
indicated by the grey triangle.}
\end{figure}

%\section{Comparison to Theory}
To shed light on the measured phase separation, we present the
predictions for the phase behavior from P-L theory \cite{Varga03}
in Fig. \ref{4theory.fig}. Four distinct types of phase diagrams
(indicated by the Greek symbols) are predicted for a mixture of
rod-like particles of equal length as a function of the diameter
ratio and the length of the thick rods. The qualitative evolution
of the phase behavior as a function of $d$ for
$L/D_{\mbox{\scriptsize thick}}=24$, which corresponds to the
experimental aspect ratio of the thick rods, is displayed in Fig.
\ref{4theory.fig}b-d. The region of I-I coexistence ($\delta$) is
not observed experimentally and is not discussed here. For small
$d$, corresponding to region $\alpha$, a transition from an
isotropic to homogenous nematic phase is predicted at low
concentrations and a N-N coexistence region bounded by a lower
critical point is predicted at high concentrations (Fig.
\ref{4theory.fig}b). Region $\alpha$ confirms that an I-N-N
coexistence region is not required for the existence of a region
of N-N coexistence, as experimentally observed (Fig.
\ref{4diagrams.fig}b). As $d$ increases within region $\alpha$ the
N-N lower critical point moves to lower rod concentrations. Region
$\beta$ is characterized by a phase diagram which exhibits an
I-N-N coexistence region capped by a region of N-N coexistence
bounded by an upper critical point, with an additional region of
N-N coexistence bounded by a lower critical point (Fig.
\ref{4theory.fig}c). Upon increasing $d$, these two regions of N-N
coexistence coalesce to form a single N-N coexistence region
(region $\gamma$, Fig. \ref{4theory.fig}d).

The phase diagram for region $\gamma$ is qualitatively similar to
that predicted by SVT \cite{vanRoij98} for $d>4.3$, and is
experimentally observed for $d\geq 3$ (Fig. \ref{4diagrams.fig}a).
The phase diagrams characterizing regions $\alpha$ and $\beta$,
however, are qualitatively different from those predicted in the
SVT ($L/D\rightarrow\infty$) \cite{vanRoij98} because they predict
a region of N-N coexistence bounded by a lower critical point.
Only an N-N coexistence region bounded by an upper critical point
is predicted in the SVT for $3.8<d<4.3$ \cite{vanRoij98}. However,
it is precisely this N-N coexistence region bounded by a lower
critical point which is measured in the {\it fd}+{\it fd}-PEG
solutions (Fig. \ref{4diagrams.fig}b) for $3>d>2$. Simply by
extrapolating between the Onsager limit ($L/D\rightarrow\infty$)
and the Carnahan-Starling hard sphere limit ($L/D\rightarrow 1$)
\cite{Parsons79}, the P-L scaling \cite{Varga03} qualitatively
reproduces the experimental phase diagram for $3>d>2$ while SVT
alone does not.

Comparing the evolution of experimental phase behavior with $d$
for our long semiflexible rods with the P-L predictions we find
that it qualitatively follows the phase behavior predicted for
short rigid rods, $L/D_{\mbox{\scriptsize thick}}\lesssim 7$. In
this case, region $\beta$ is bypassed and the N-N region bounded
by the lower critical point coalesces directly with the I-N region
creating an I-N-N coexistence region. It has been shown in
simulations that the excluded volume of a flexible rod is
equivalent to the excluded volume of a shorter but thicker rigid
rod \cite{Fynewever98}. Thus we expect long flexible rods to
exhibit a phase behavior similar to that predicted for shorter
rigid rods, as observed. Additionally, we observe that the
experimental I-N-N coexistence is stable to much lower diameter
ratios, $d\sim 3$, than predicted. We suggest that this is because
the thin-thick rod interactions are non-additive. Because
$D_{\mbox{\scriptsize thin}}$ is defined by electrostatic
repulsion, nothing prevents the polymer coated virus (thick rods)
from touching the bare surface of the {\it fd} (thin rods).
Subsequently, the thin-rod interaction diameter is effectively
smaller ($d$ is effectively larger). Theoretically, one of the
challenges that remains is to incorporate non-additivity and
flexibility into theories for the binary rod phase behavior.

%\section{Conclusions}
In conclusion, we have studied the phase behavior of the first
experimental system of thick and thin rods. Onsager's SVT
qualitatively reproduces the main features seen in our
experimental binary rod mixtures at large diameter ratios, but
does not accurately capture the evolution of the phase behavior
from a totally miscible nematic to a demixed nematic-nematic
state. By incorporating the higher virial coefficients in an
approximate fashion with the Parsons-Lee scaling of the free
energy for hard rods \cite{Varga03}, we proposed an evolution of
the binary rod phase diagram with diameter ratio which
qualitatively describes the experimental findings. Our
experimental and theoretical results show that an I-N-N
coexistence region is not required for the existence of a region
of N-N coexistence in contrast to past predictions. However, the
N-N upper critical point, which is predicted for very long rods in
both the SVT and Parsons-Lee theory has not yet been observed
experimentally; further experimental or computational studies of
binary mixtures of longer, more rigid rods, may reveal this upper
critical point.
%We can now systematically vary diameter and length ratios of the
%virus rods from $1<d<5$ and $1<l<3$, respectively, inviting future
%studies into the presence of an N-N region bounded by an upper
%critical point in binary mixtures, as well as investigate
%tri-disperse and quantitatively polydisperse mixtures of rods.

\begin{acknowledgments}
The experimental research (KP and SF) was supported by an NSF/DMR
grant. The theoretical work (AG and GJ) was supported by the
Engineering and Physical Sciences Research Council, the Joint
Research Equipment Initiative, and the Royal Society-Wolfson
Foundation. We acknowledge Rene van Roij for helpful discussions
and unpublished theoretical results.
\end{acknowledgments}

\end{document}